\documentclass[letter,apj]{emulateapj-rtx4}
\usepackage{graphicx}
\usepackage{enumerate}
\usepackage{amssymb, amsmath}
\usepackage{mathtools}
\usepackage{natbib}
\usepackage{color}
\usepackage{hyperref}
\usepackage{ulem}
\usepackage{soul}
\newcommand{\ias}{1}
\newcommand{\elte}{2}
\newcommand{\sagan}{3}

\begin{document}

\title{Disrupted Globular Clusters Can Explain the Galactic Center Gamma Ray Excess}
\author{
Timothy D.~Brandt\altaffilmark{\ias,\sagan} and
Bence Kocsis\altaffilmark{\ias,\elte}
}

\altaffiltext{\ias}{Institute for Advanced Study, Einstein Dr., Princeton, NJ}
\altaffiltext{\elte}{E\"otv\"os University, P\'azm\'any P. s. 1/A, Budapest, Hungary}
\altaffiltext{\sagan}{NASA Sagan Fellow}

\begin{abstract}
The {\it Fermi} satellite has recently detected gamma ray emission from the central regions of our Galaxy.  This may be evidence for dark matter particles, a major component of the standard cosmological model, annihilating to produce high-energy photons.  We show that the observed signal may instead be generated by millisecond pulsars that formed in dense star clusters in the Galactic halo.  Most of these clusters were ultimately disrupted by evaporation and gravitational tides, contributing to a spherical bulge of stars and stellar remnants. The gamma ray amplitude, angular distribution, and spectral signatures of this source may be predicted without free parameters, and are in remarkable agreement with the observations. These gamma rays are from fossil remains of dispersed clusters, telling the history of the Galactic bulge.
\end{abstract}

\maketitle

\section{Introduction} \label{sec:intro}

While there are strong indications for the existence of cold dark matter from its gravitational effects \citep[e.g.][]{Planck_2014}, there has not yet been any conclusive direct or indirect detection of the corresponding dark matter particles.  One promising avenue to look for these particles is through annihilation in which two dark matter particles (a particle and its antiparticle) convert into high energy photons that we can observe. The dark matter annihilation signal is expected to be strongest where the density of dark matter is highest, i.e., in the centers of galaxies.  

Detailed analyses of the {\it Fermi} satellite's map of the gamma-ray sky have revealed an excess around the Galactic center peaking at energies of $\sim$2 GeV \citep[e.g.][]{Hooper+Goodenough_2011, Gordon+Macias_2013, Daylan+Finkbeiner+Hooper+etal_2014}.  This excess appears to be roughly spherical and extends at least $\sim$10--20$^\circ$ (1.5--3 kpc) from Sgr A*, the Galaxy's central supermassive black hole.  
Remarkably, this signal can be interpreted as photons from annihilating $\sim$30 GeV dark matter particles \citep{Hooper+Goodenough_2011, Daylan+Finkbeiner+Hooper+etal_2014}. In order to confirm this extraordinary interpretation, one must carefully rule out all other astrophysical sources. Possible alternatives include millisecond pulsars (MSPs), rapidly spinning neutron stars that are observed in other regions of the Galaxy with very similar gamma ray spectra to that of the observed excess \citep{Gordon+Macias_2013, Abazajian+Kaplinghat_2012, Yuan+Zhang_2014, Abazajian_2011J, Mirabal_2013, Yuan+Ioka_2015, Petrovic+Serpico+Zaharijas_2015}; highly magnetized young pulsars created in the innermost nuclear star cluster \citep{OLeary_2015}; injection of cosmic-ray protons
\citep{Carlson+Profumo_2014}; or cosmic ray outbursts \citep{Petrovic+Serpico+Zaharija_2014}. However, it remains to be shown that any of these sources is sufficiently abundant and spatially extended to explain the gamma-ray excess. 

Energetic photons have also been observed from within the central few pc around Sgr A* itself, extending from soft X-rays to $\sim$100 TeV gamma rays \citep{Baganoff+Bautz+Brandt+etal_2001,Aharonian+Akhperjanian+Aye+etal_2004,Belanger+Goldwurm+Renaud+etal_2006,Perez+Hailey+Bauer+etal_2015}. The origin of this emission is subject to debate; see \cite{vanEldik_2015} for a review.  The region near the event horizon of Sgr A* is likely responsible for bright outbursts in soft X-rays \citep{Baganoff+Bautz+Brandt+etal_2001}, but this scenario struggles to explain the steady emission at much higher energies.  Alternative explanations for the GeV and TeV flux include the supernova remnant Sgr A East \citep{Crocker+Fatuzzo+Jokipii+etal_2005}, though this is strongly disfavored based on its observed offset from the very high energy emission centered on Sgr A* \citep{Acero+Aharonian+Akhperjanian+etal_2010}.  Secondary emission from particles accelerated by Sgr A* is another candidate, either in a steady state or from a past burst of accretion \citep[e.g.][]{Atoyan+Dermer_2004,Aharonian+Neronov_2005,Chernyakova+Malyshev+Aharonian+etal_2011}.  Most of these scenarios cannot account for both the GeV and TeV emission.  In contrast, a population of $\sim$1000 MSPs in the inner few pc could account for the emission from GeV through 100 TeV \citep{Bednarek+Sobczak_2013}.  None of these scenarios seek to explain the GeV excess extending several kpc from Sgr A*.  

The pulsar population in the Galactic center has long been sought to test the theory of gravity \citep[and references therein]{Pfahl+Loeb_2004,Liu+etal_2012} and the existence of intermediate mass black holes and gravitational waves \citep{Kocsis+Ray+PortegiesZwart_2012}. 
Extended multiwavelength observations were conducted which should have detected a significant fraction of the most common second-period pulsars, but only four were seen. This missing pulsar problem indicates that the formation and/or retention of ordinary pulsars may be inefficient in this region \citep{Dexter+OLeary_2014,Macquart+Kanekar_2015}. However, these searches did not significantly constrain the number of MSPs, especially at the relatively large galactocentric distances of 0.1--1 kpc where the gamma ray excess is observed.

In this paper, we argue that the MSPs needed to produce the gamma ray excess were not made under the present conditions of the Galactic bulge, but were produced in dense globular clusters that have since dissolved.  The population of globular clusters constitutes a key component in the theory of galaxy evolution, the formation of galactic bulges, and nuclear star clusters \citep{Tremaine+Ostriker+Spitzer_1975,Arca-Sedda+Capuzzo-Dolcetta_2014,Capuzzo-Dolcetta1993,Lotz+etal_2001,Bekki+etal_2004,Capuzzo-Dolcetta+Miocchi_2008,Capuzzo-Dolcetta+Mastrobuono-Battisti_2009,Agarwal+Milosavljevic_2011,Hartmann+etal_2011,Leigh+Boker+Knigge_2012,Antonini+Capuzzo-Dolcetta+Mastrobuono-Battisti_2012,Antonini_2013,Antonini_2014,denBrok+etal_2014,Kruijssen_2014,Perets+Mastrobuono-Battisti_2014,Antonini+Barausse+Silk_2015a,Antonini+Barausse+Silk_2015b}.  The clusters we see today are the ones that have survived throughout the evolution of the Galaxy, and may be a small fraction of the initial cluster population.  

We organize the paper as follows.  In Section \ref{sec:msps_clusters}, we discuss the relationship of MSPs to globular clusters and the evolution of the Galaxy's population of globular clusters.  Section \ref{sec:lumscale} discusses the scaling of the gamma ray luminosity to the predicted population of disrupted clusters, while Section \ref{sec:pred_excess} presents the predictions of this model for the {\it Fermi} excess.  Sections \ref{sec:maxlum} and \ref{sec:MSPspec} discuss two objections to a MSP explanation of the excess, the high end of the luminosity function and the average spectrum.  We discuss prospects for radio detections of our predicted MSPs in Section \ref{sec:discussion} and conclude with Section \ref{sec:conclusions}.

\section{Millisecond Pulsars from Disrupted Globular Clusters} \label{sec:msps_clusters}

MSPs are thought to be ``recycled'' pulsars, spun up by the accretion of material from a close binary companion \citep{Bhattacharya+vandenHeuvel_1991}.  These close binaries are formed and driven to smaller separations in dense stellar environments where the rate of stellar dynamical encounters is high.  Once these interactions have sufficiently decreased the binary separation, the neutron star's companion 
transfers material and angular momentum, and reduces the neutron star's magnetic field.  This phase lasts $\sim$10$^7$--10$^9$ years and is visible, for low-mass companions, as a low mass X-ray binary (LMXB) \citep{Ivanova+etal_2008}.  A long-lived MSP remains after the mass transfer stops. While the strong magnetic braking in ordinary pulsars leads to a rapid spindown, MSPs persist for $\sim$10$^{10}$ years \citep{Bhattacharya+vandenHeuvel_1991}; these pulsars can long outlive their birth clusters.

The highest abundances of MSPs are found in globular clusters, the old dense stellar islands orbiting in the galactic halo. There are several indications that a fraction of the stellar mass of galactic bulges may have been formed by dissolving globular clusters \citep{Tremaine+Ostriker+Spitzer_1975,Arca-Sedda+Capuzzo-Dolcetta_2014,Gnedin+Ostriker+Tremaine_2014}. The distribution of old globular clusters within galaxies and the geometry of galactic bulges are both approximately spherical. The number density of globular clusters increases inwards on kpc scales, but shows a relative decrease within the galactic bulge. A tight correlation is observed between the mass of the galactic bulge and the number of globular clusters \citep{Harris+Poole+Harris_2014}. 

Massive globular clusters spiral in towards the Galactic center due to dynamical friction.  In the central kpc, the tidal gravitational field of the Galaxy may exceed the attractive field of the cluster stars, stripping the cluster from its outskirts and eventually down to its core.  The cluster then spills its entire contents, including the MSPs in its core, into a spherical shell about the Galactic center.  Since MSPs are long-lived, they remain bright in gamma rays after the cluster is disrupted. The high dynamical encounter rates needed to form new LMXBs and new MSPs are, however, strongly suppressed after the disruption of the globular cluster. Therefore, the MSP population will be frozen at the time and orbit of the cluster's disruption while LMXBs (precursors to MSPs) will burn out within $\sim$10$^8$ years.  As a result, the ratio of LMXBs to MSPs from disrupted globular clusters becomes much lower in the galactic bulge than in the surviving globular clusters we observe today. Indeed, LMXBs are observed to be rare in the bulge \citep{Revnivtsev+Lutovinov+Churazov+etal_2008, Cholis+Hooper+Linden_2015}.

We model the distribution of globular clusters and the Galactic bulge following \cite{Gnedin+Ostriker+Tremaine_2014}, who account for mass loss from passive stellar evolution, cluster evaporation, infall due to dynamical friction, and tidal disruption; we adopt all of their fiducial parameters. This simple model was set up to reproduce the radial and mass distribution of extant globular clusters in the halo with no eye towards reproducing the {\it Fermi} excess. The distribution of mass from dissolved globular clusters is shown in Figure 3 of \cite{Gnedin+Ostriker+Tremaine_2014}; we use this result directly.  It is a cored radial profile with $\rho(r)\sim r^{-2.2}$ and an enclosed mass $\sim$10$^8$ $M_\odot$ at 1 kpc. We assume that most gamma-ray sources in globular clusters formed sufficiently quickly that the luminosity per unit mass in a population of disrupted clusters may be approximated by the observed value in surviving globular clusters. Their radial distribution is set by their orbits at their points of disruption in the model. We do not tune the fiducial model of \cite{Gnedin+Ostriker+Tremaine_2014}; our approach has no free parameters. 

The Appendix gives a brief overview of the \cite{Gnedin+Ostriker+Tremaine_2014} model and calculations, with equations giving the characteristic disruption timescales.  We refer the reader to that paper for a more thorough discussion.

\section{Scaling the Gamma Ray Luminosity} \label{sec:lumscale}

We compute the gamma ray luminosity per unit stellar mass for the globular clusters studied by \cite{Abdo+Ackerman+Ajello+etal_2010} and listed in Table \ref{tab:clusters}.  Of these eleven clusters, \cite{Abdo+Ackerman+Ajello+etal_2010} reported gamma ray detections for eight.  We use the more recent 2 GeV fluxes measured by \cite{Cholis+Hooper+Linden_2015}; this data set includes fluxes for two of the three globular clusters that were previously undetected.  We adopt the cluster distances compiled by \cite{Abdo+Ackerman+Ajello+etal_2010} and convert the absolute $V$ magnitudes of \citealt{Harris_1996} (2010 edition) to mass by assuming a mass-to-light ratio of 3 $M_\odot$/$L_\odot$, appropriate for an old, slightly metal-poor population \citep{Maraston_2005}.  The mass of Terzan 5 is uncertain due to its very large extinction \citep{Lanzoni+Ferraro+Dalessandro+etal_2010}, with those authors favoring a value $2 \times 10^6$ $M_\odot$ compared to the $3 \times 10^5$ $M_\odot$ implied by the absolute $V$ magnitude given by \cite{Harris_1996}, but this has little effect on our results.

\begin{deluxetable}{lcccr}
\tablewidth{0pt}
\tablecaption{Properties of Globular Clusters}
\tablehead{
    Name & 
    Dist$_{\rm A10}$\tablenotemark{a} &
    Dist$_{\rm H10}$\tablenotemark{b} &
    $M_V$\tablenotemark{b} &
    $F_{2\,\rm GeV}$\tablenotemark{c} \\
    &
    (kpc) &
    (kpc) &
    (mag) &
    }
\startdata
47 Tuc & $4.0\pm 0.4$ & 4.5 & $-9.42$ & 5.6 \\
$\omega$ Cen & $4.8 \pm 0.3$ & 5.2 & $-10.26$ & 2.8 \\
M 62 & $6.6 \pm 0.5$ & 6.8 & $-9.18$ & 3.8 \\
NGC 6388 & $11.6 \pm 2.0$ & 9.9 & $-9.41$ & 3.4 \\
Terzan 5 & $5.5 \pm 0.9$ & 6.9 & $-7.42$\tablenotemark{*} & 12.6 \\
NGC 6440 & $8.5 \pm 0.4$ & 8.5 & $-8.75$ & 2.9 \\
M 28 & $5.1 \pm 0.5$ & 5.5 & $-8.16$ & 3.8 \\
NGC 6652 & $9.0 \pm 0.9$ & 10.0 & $-6.66$ & 1.4 \\
NGC 6541 & $6.9 \pm 0.7$ & 7.5 & $-8.52$ & 0.9 \\
NGC 6752 & $4.4 \pm 0.1$ & 4.0 & $-7.73$ & 0.5 \\
M 15 & $10.3 \pm 0.4$ & 10.4 & $-9.19$ & \ldots
\enddata
\tablenotetext{a}{From \cite{Abdo+Ackerman+Ajello+etal_2010}, references therein}
\tablenotetext{b}{From \citealt{Harris_1996} (2010 edition)}
\tablenotetext{c}{$E^2 dN/dE$, units are $10^{-9}$ GeV\,cm$^{-2}$\,s$^{-1}$. From \cite{Cholis+Hooper+Linden_2014}.}
\tablenotetext{*}{Uncertain due to very high extinction}
\label{tab:clusters}
\end{deluxetable}

Our approach gives a 2 GeV flux density at a distance of 8.3 kpc of $2 \times 10^{-15}$ GeV\,cm$^{-2}$\,s$^{-1}$\,$M_\odot^{-1}$.  The variance on this value as estimated from bootstrap resampling is 30\%, with additional uncertainties from the cluster luminosities and distances, and possible systematic variations of cluster properties with galactocentric radius.  Increasing the mass of Terzan 5 to $2 \times 10^6$ would decrease this value by $\sim$10\%, while adopting the \citealt{Harris_1996} (2010 edition) distances would increase it by a similar factor.  There are also uncertainties about how representative these extant clusters are of the initial population and indeed about the initial population itself; we therefore (somewhat arbitrarily) adopt a factor of 2 (0.3 dex) as the uncertainty in our gamma ray luminosity scaling.  

We neglect systematic variations in the number of MSPs per unit globular cluster mass as a function of the cluster properties and evolutionary stage. The formation rate and the total number of MSPs are observed to correlate with the rate of encounters, not simply the cluster mass \citep{Hui+Cheng+Taam_2010, Bahramian+etal_2013}, while the encounter rates are strongly affected by core collapse and the primordial binary fraction which are not well understood \citep{Binney+Tremaine_2008}. Indeed, the clusters listed in Table \ref{tab:clusters} have only a weak correlation between stellar mass and gamma ray luminosity. Future studies are needed to examine more detailed models of the MSPs within a population of globular clusters.

\section{The Predicted {\it Fermi} Excess} \label{sec:pred_excess}

\begin{figure}
\centering
\includegraphics[width=\linewidth]{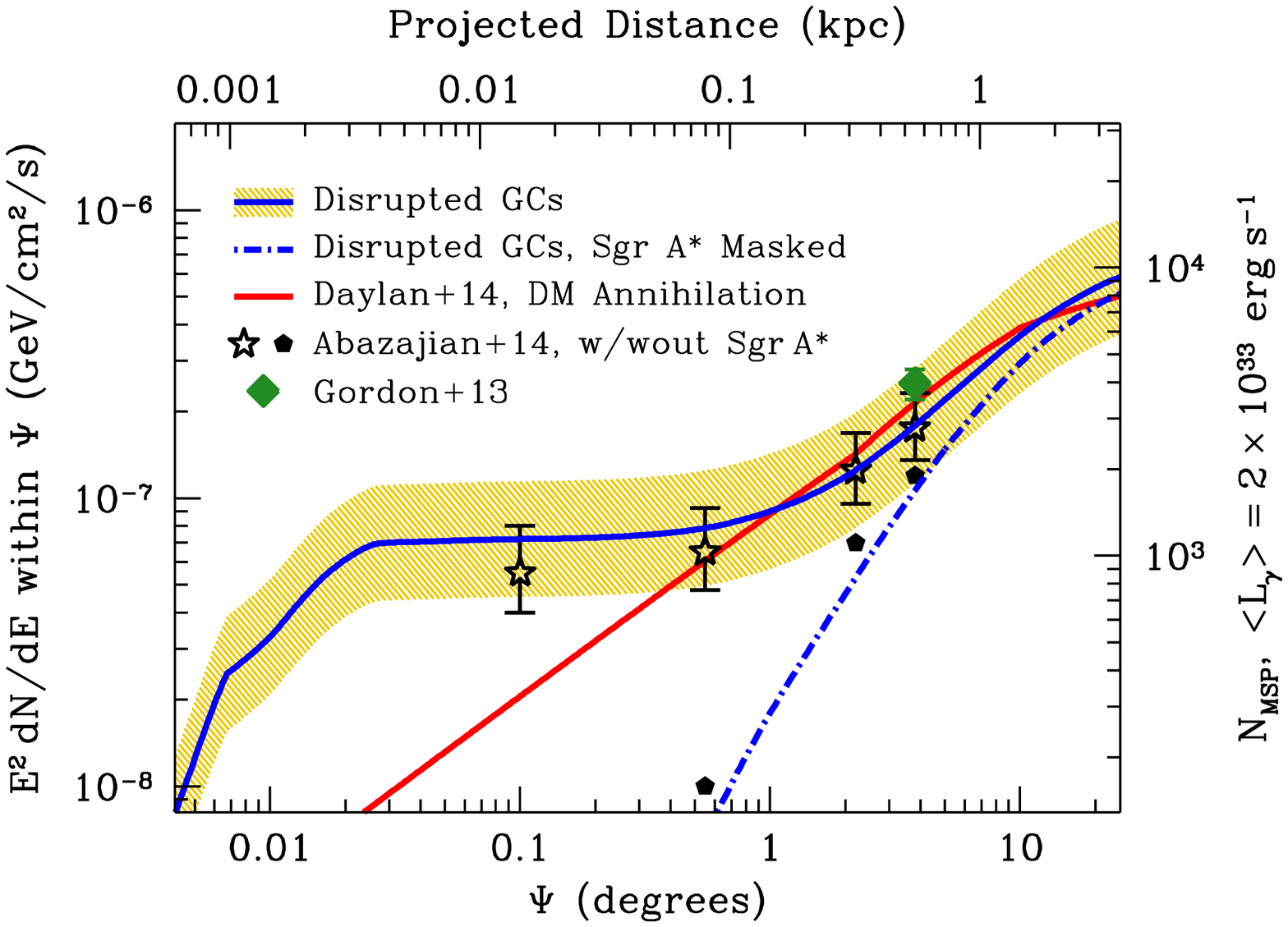}
\caption{
Integrated flux (within an angle $\Psi$ of the Galactic center) of the {\it Fermi} excess at 2 GeV by \cite{Gordon+Macias_2013}, \cite{Daylan+Finkbeiner+Hooper+etal_2014} and \cite{Abazajian+Canac+Horiuchi+etal_2014}, compared to the prediction (solid blue curve) from disrupted globular clusters, assuming the same gamma ray luminosity per unit mass as for intact clusters.  The yellow hatching shows a factor of two uncertainty; the right axis shows the approximate number of enclosed MSPs.  
The gamma ray signal from scaling the disrupted globular clusters of \cite{Gnedin+Ostriker+Tremaine_2014} correctly predicts all measurements, including an unresolved source around Sgr A* seen by \cite{Abazajian+Canac+Horiuchi+etal_2014}, with no free parameters.  The black open stars include Sgr A* and the filled pentagons exclude it; we have also shown the disrupted globular cluster prediction with the inner $0.\!\!^{\circ}1$ masked for comparison (blue dotted-dashed curve).  We interpret this unresolved source as emission from MSPs in a nuclear star cluster.  \cite{Daylan+Finkbeiner+Hooper+etal_2014} and \cite{Gordon+Macias_2013} include this unresolved source in their diffuse fits.  
}
\label{fig:integral_GCE}
\end{figure}

\begin{figure}
\centering
\includegraphics[width=\linewidth]{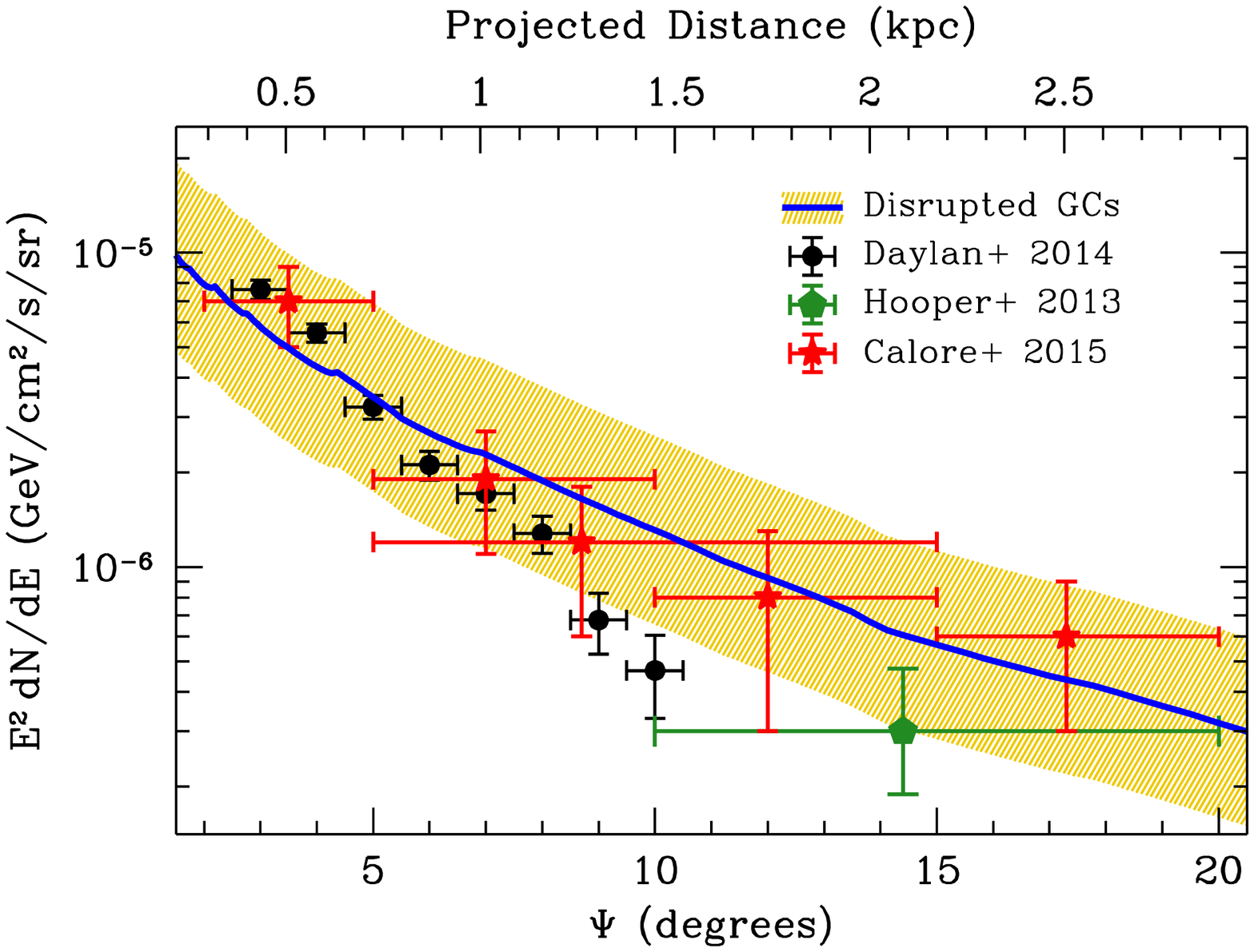}
\caption{Differential measurements of the Galactic Center excess at 2 GeV by \cite{Hooper+Slatyer_2013}, \cite{Daylan+Finkbeiner+Hooper+etal_2014}, and \cite{Calore+Cholis+Weniger_2015}, compared to the prediction from scaling the \cite{Gnedin+Ostriker+Tremaine_2014} disrupted globular clusters to the same gamma ray luminosity per unit mass as for intact clusters.  We have included a factor of two uncertainty (yellow hatching) on the globular cluster prediction (blue curve).  The blue curve and yellow hatching are not fitted to the data; they include no free parameters. }
\label{fig:differential_GCE}
\end{figure}

With an independent model of the population of disrupted globular clusters \citep{Gnedin+Ostriker+Tremaine_2014} and a scaling of total globular cluster mass to gamma ray luminosity (Section \ref{sec:lumscale}), we may compute the predicted {\it Fermi} GeV signal.  Our results for the integrated flux within a circular region around the Galactic center, and the approximate number of enclosed MSPs, are shown in Figure \ref{fig:integral_GCE} as a function of the angular distance to the center.  Figure \ref{fig:differential_GCE} shows the differential flux within circular annuli. Each figure shows the recent measurements of the {\it Fermi} excess to be in excellent agreement with our predictions.  

The integral flux (Figure \ref{fig:integral_GCE}) contains two particularly noteworthy results.  First, the red line in Figure \ref{fig:integral_GCE} is a discontinous set of dark matter profiles fit at different annuli: the best annihilation fit to the data is not a physically meaningful dark matter profile.  At separations $>$$0.\!\!^{\circ}5$, which account for nearly all of the flux in the \cite{Daylan+Finkbeiner+Hooper+etal_2014} fits, disrupted globular clusters provide a physical motivation for the form of this signal.  Second, \cite{Abazajian+Canac+Horiuchi+etal_2014} fit for an unresolved source at the Galactic center, Sgr A*, which can be explained as MSP emission in a nuclear star cluster.  We have added this central source to the diffuse emission fitted by \cite{Abazajian+Canac+Horiuchi+etal_2014} to obtain the open black stars in Figure \ref{fig:integral_GCE}.  We have placed the leftmost black star at $0.\!\!^{\circ}1$, roughly the angular resolution of the {\it Fermi} telescope.  The actual extent of the nuclear star cluster is predicted to be just a few pc, or $\sim$$0.\!\!^{\circ}02$.  To facilitate a direct comparison, we have also removed emission from the inner $0.\!\!^{\circ}1$ of the disrupted cluster prediction, and indicated the resulting diffuse gamma ray signal with a dotted-dashed blue line.  Our prediction matches the gamma-ray flux at all separations.

The existence of strong gamma ray emission from a nuclear star cluster reported by \cite{Abazajian+Canac+Horiuchi+etal_2014} further supports the disrupted globular cluster hypothesis. \cite{Faucher-Giguere+Loeb_2011} found that the total encounter rate in the inner parsec of the Galaxy is similar to that of the globular cluster Terzan 5, so the formation rate of MSPs (and resulting gamma ray luminosity) may be similar. The 2 GeV flux of Terzan 5 would be just $5 \times 10^{-9}$ GeV\,cm$^{-2}$\,s$^{-1}$ at 8.3 kpc. Unless the gravitational potential of Sgr A* retains a much higher number of neutron stars than in globular clusters, this scenario can only explain $\sim$10\% of the flux seen by \cite{Abazajian+Canac+Horiuchi+etal_2014} and shown in Figure \ref{fig:integral_GCE}.  MSPs deposited in the nuclear star cluster by massive globular clusters were also suggested by \cite{Bednarek+Sobczak_2013} as an explanation of the observed TeV photons from around Sgr A*.  Those authors required a population of $\sim$1000--3000 MSPs, fully consistent with our results both in the Galactic center and at larger separations.  Recent 20--40 keV X-ray observations by the {\it NuSTAR} satellite can also be explained by a large population of MSPs in the inner few pc \citep{Perez+Hailey+Bauer+etal_2015}.

\section{The Maximum Luminosity of MSPs} \label{sec:maxlum}

One objection to a population of bulge MSPs as the source of the {\it Fermi} excess is the paucity of individually identified high luminosity gamma ray pulsars detected as point sources within $\sim$10$^\circ$ of the Galactic center \citep{Cholis+Hooper+Linden_2015}. \cite{Lee+Lisanti+Safdi+etal_2015} found evidence that the GeV excess is from unresolved point sources with a 1.9--12 GeV flux cutoff at $\sim$1.5--$2 \times 10^{-10}$ photons\,cm$^{-2}$\,s$^{-1}$, for a 0.1--100 GeV luminosity of $\sim$1.5--$2.5\times 10^{34}$ erg\,s$^{-1}$ at 8.3 kpc.  \cite{Cholis+Hooper+Linden_2014} found a very hard luminosity function for MSPs, with most of the luminosity contributed by objects above a few $10^{34}$ erg\,s$^{-1}$, which should have been detected as {\it Fermi} point sources.  We reanalyze the data of \cite{Cholis+Hooper+Linden_2014} and reexamine the cutoff at high luminosities.  

\cite{Cholis+Hooper+Linden_2014} use a sample of 59 field MSPs to determine the luminosity function (listed in their Table IV).  The distances they adopt are taken from the ATNF pulsar database \citep{Manchester+Hobbs+Teoh+etal_2005}, available at \url{http://www.atnf.csiro.au/people/pulsar/psrcat/}, which uses the model Galaxy of \cite{Taylor+Cordes_1993}, hereafter TC93, to convert observed dispersion measures into distances.  The {\it Fermi} Second Pulsar Catalog (2PC) distances \citep{Abdo+Ajello+Allafort+etal_2013} listed in the same table generally use the same dispersion measures to calculate distances, but rely on the newer and more accurate NE2001 model of the Galaxy \citep{Cordes+Lazio_2002}.  The newer distances are systematically lower, particularly out of the Galactic plane.  Unlike the older TC93 model, the NE2001 model supplies enough free electrons to account for the observed dispersion measures of nearly all Galactic pulsars \citep{Cordes+Lazio_2002}.  \cite{Cholis+Hooper+Linden_2014} only used the 44 MSPs with $|b|>10^\circ$ to determine the luminosity function; the revised distances thus have a large effect on the results.  

\begin{deluxetable}{lcccr}
\tablewidth{0pt}
\tablecaption{Distances to Field Millisecond Pulsars}
\tablehead{
    Name & 
    DM &
    $D_{\rm TC93}$ &
    $D_{\rm NE2001}$ & 
    Ref\tablenotemark{a} \\
    &
    pc\,cm$^{-3}$ &
    (kpc) &
    (kpc) &
    }
\startdata
J0307$+$7443 &  6.4 & 0.34 & 0.6 & R12 \\
J0533$+$6759 & 57.4 & 6.66 & 2.4 & R12 \\
J0605$+$3757 & 21.0 & 1.16 & 0.7 & R12 \\
J1137$+$7528 & 29.2 & 19.53 & 1.5 & \tablenotemark{*} \\
J1142$+$0119 & 19.2 & 2.04 & 0.9 & R12 \\
J1301$+$0833 & 13.2 & 0.91 & 0.7 & R12 \\
J1302$-$3258 & 26.2 & 1.86 & 1.0 & R12 \\
J1311$-$3430 & 37.8 & 3.72 & 1.4 & R13 \\
J1312$+$0051 & 15.3 & 1.15 & 0.8 & R12 \\
J1543$-$5149 & 50.9 & 1.46 & 2.4 & N14 \\
J1544$+$4936 & 23.2 & 2.30 & 1.2 & R12 \\
J1630$+$3734 & 14.1 & 0.85 & 0.9 & R12 \\
J1640$+$2224 & 18.4 & 1.15 & 1.16 & L05 \\
J1732$-$5049 & 56.8 & 1.81 & 1.3 & V09 \\
J1745$+$1017 & 23.9 & 1.36 & 1.3 & R12 \\
J1811$-$2405 & 60.6 & 1.70 & 1.8 & N14 \\
J1816$+$4510 & 38.9 & 4.20 & 2.4 & R12 \\
J1843$-$1113 & 60.0 & 1.97 & 1.7 & H04 \\
J2129$-$0429 & 16.9 & 1.03 & 0.9 & R12 \\
J2256$-$1024 & 14   & 0.91 & 0.65 & B13
\enddata
\tablenotetext{a}{References abbreviated as: B13 \citep{Breton+vanKerwijk+Roberts+etal_2013}; C14 \citep{Cholis+Hooper+Linden_2014}; H04 \citep{Hobbs+Faulkner+Stairs+etal_2004}; L05 \citep{Lohmer+Lewandowski+Wolszczan+etal_2005}; N14 \citep{Ng+Bailes+Bates+etal_2014}; R12 \citep{Ray+Abdo+Parent+etal_2012}; R13 \citep{Ray+Ransom+Cheung+etal_2013}; V09 \citep{Verbiest+Bailes+Coles+etal_2009}.  }
\tablenotetext{*}{\url{http://astro.phys.wvu.edu/GalacticMSPs/GalacticMSPs.txt}, see text for details}
\label{tab:field_msp}
\end{deluxetable}

Table \ref{tab:field_msp} lists the NE2001 distances for all pulsars without 2PC distances listed in \cite{Cholis+Hooper+Linden_2014}.  In the case of J1843$-$1113, \cite{Hobbs+Faulkner+Stairs+etal_2004} have provided a distance using the TC93 model Galaxy.  We have taken their dispersion measure and converted it into the tabulated NE2001 distance estimate.  One MSP, J1137+7528, does not appear in any database.  The only reference to this pulsar that we were able to find (apart from the {\it Fermi} Third Point Source catalog, \citealt{Acero+Ackermann+Ajello+etal_2015}) was in a table of pulsar data available at \url{http://astro.phys.wvu.edu/GalacticMSPs/GalacticMSPs.txt}.  This site lists a dispersion measure of 29.2 pc cm$^{-3}$ which, according to the NE2001 free electron model, implies a distance of 1.5 kpc.  

We also note that these new distances have significant uncertainties.  The brightest MSP in our sample, for example, is J0614$-$3329, which was identified with {\it Fermi}  by \cite{Ransom+Ray+Camilo+etal_2011}.  This is a generally unremarkable MSP apart from the very high gamma ray efficiency (greater than unity) implied by its dispersion measure distance of 1.9 kpc.  As \cite{Ransom+Ray+Camilo+etal_2011} note, J0614 is nearly tangent to the Gum nebula where the NE2001 model has a very steep gradient in dispersion measure (and hence, in derived distance); those authors suggest that the true distance is likely to be closer by a factor of 2 or more.  Decreasing its distance by such a factor would reduce its integrated gamma ray luminosity to $\sim$$10^{34}$ erg\,s$^{-1}$ and remove the need to invoke a very large gamma ray efficiency and/or beaming factor.

Other pulsars, including those just below the cutoff, could also have distance errors.  Random errors, however, will tend to smear out a distribution; a deconvolution should sharpen the cutoff.  Also, an anomalously large distance (by a factor of 2, say) requires simply an unmodeled clump of ionized gas along the line-of-sight.  An anomalously small distance requires a void or bubble; the required void for the same fractional distance error becomes larger with increasing distance.  A factor of 2 error in a 1 kpc distance would require a 1 kpc completely empty void (to go along with the 1 kpc of properly modeled free electron density).

\begin{figure}
\centering
\includegraphics[width=\linewidth]{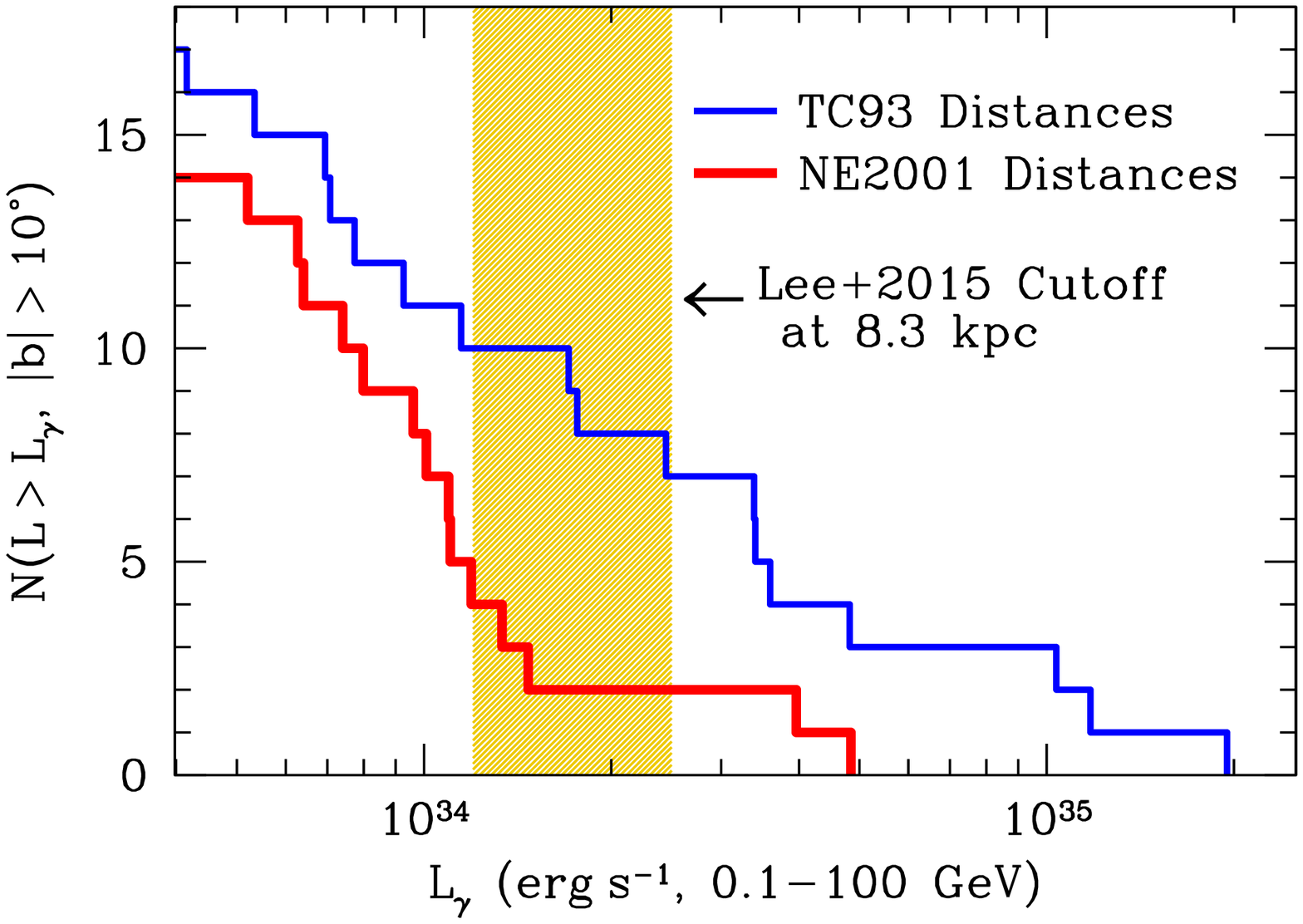}
\caption{Cumulative number of MSPs in the \cite{Cholis+Hooper+Linden_2015} sample with $|b|>10^\circ$ adopting the \cite{Taylor+Cordes_1993} (TC93, blue line) and an updated model \citep[NE2001, red line]{Cordes+Lazio_2002} of the Galaxy's dispersion measure. The latter leaves just two MSPs sufficiently luminous to detect as point sources near the Galactic center.
One, J0614$-$3329 is likely to be at least a factor of 4 less luminous than shown \citep{Ransom+Ray+Camilo+etal_2011}; the other, J0218$+$4232, is just marginally ($\sim$1$\sigma$) more luminous than the Galactic center detection threshold \citep{Abdo+Ajello+Allafort+etal_2013}. The vertical shading shows the luminosity cutoff needed to reproduce the gamma ray excess with an unresolved source population \citep{Lee+Lisanti+Safdi+etal_2015}; \cite{Bartels+Krishnamurthy+Weniger_2015} favor a somewhat higher cutoff.
}
\label{fig:lum_cutoff}
\end{figure}

Figure \ref{fig:lum_cutoff} shows the cutoff at high luminosities in the \cite{Cholis+Hooper+Linden_2015} MSP sample using both the TC93 and the NE2001 distances.  For reference, we have also shown the approximate luminosity cutoff favored by \cite{Lee+Lisanti+Safdi+etal_2015} of $\sim$1.5--2 photons\,cm$^{-2}$\,s$^{-1}$ between 1.9 and 12 GeV, transformed into 0.1--100 GeV flux assuming the well-measured spectrum of J0614$-$3329 and placed at a distance of 8.3 kpc.  There are only two MSPs above this cutoff.  One, J0614$-$3329 itself, was discussed above; \cite{Ransom+Ray+Camilo+etal_2011} argue that its true distance is likely to be at least a factor of $\sim$2 smaller than that implied by its dispersion measure.  The other is J0218$+$4232, for which \cite{Abdo+Ajello+Allafort+etal_2013} quote a factor of 2 error in the luminosity as a result of dispersion measure uncertainties in the NE2001 model, making it more luminous than $2 \times 10^{34}$ erg\,s$^{-1}$ by just $\sim$1$\sigma$.  The updated pulsar distances support the luminosity cutoff needed by \cite{Lee+Lisanti+Safdi+etal_2015} and \cite{Bartels+Krishnamurthy+Weniger_2015} to account for the GeV excess with unresolved point sources.  \cite{Bartels+Krishnamurthy+Weniger_2015} favor a cutoff at higher luminosities than \cite{Lee+Lisanti+Safdi+etal_2015} (though over a different bandpass).

\section{The Average Spectrum of MSPs} \label{sec:MSPspec}

We now turn to the spectrum of an unresolved distribution of {\it Fermi} MSPs, estimating the integrated light using the field MSPs listed in \cite{Cholis+Hooper+Linden_2014}.  The spectrum of MSPs has been suggested to differ modestly from that of the observed GeV excess \citep{Cholis+Hooper+Linden_2015}, arguing against their ability to produce the gamma rays around the Galactic center.  We use the field MSPs under the assumption that many of them have similar origins to the one we suggest for the central bulge population, in the cores of stellar clusters disrupted long ago.  We also account for the fact that {\it Fermi}'s sensitivity is a strong function of photon energy \citep{Acero+Ackermann+Ajello+etal_2015}.  Following \cite{Cholis+Hooper+Linden_2014}, we do not apply a cut in Galactic latitude $b$ (as we and they did for the high luminosity cutoff, Section \ref{sec:maxlum}).  Applying such a cut would make the spectra we show in this section slightly harder, and lessen the tension with the \cite{Daylan+Finkbeiner+Hooper+etal_2014} spectrum of the GeV excess.

The MSP sample of \cite{Cholis+Hooper+Linden_2014} is not selected at a single frequency: for a fixed 2 GeV flux density with little emission from higher energy photons, very soft sources are easier to detect than harder sources.  Only the soft sources supply enough photons at energies $<$1 GeV to contribute significantly to their detectability.  We therefore create a 2 GeV-selected sample from the \cite{Cholis+Hooper+Linden_2014} MSPs by including only those with a 1--3 GeV test statistic of $8^2$, corresponding to a signal-to-noise ratio of at least 8 in this band.  Of the 59 MSPs, 45 pass this cut.  This sample of 45 MSPs should be complete independently of spectral shape.  

Under the (unlikely) assumption that our field MSPs form a flux-limited sample of a spatially uniform population, we can derive a weighted average of the individual MSP spectra that matches the spectrum expected for a population at fixed distance.  Uniformly weighting each MSP's spectrum is equivalent to assuming a volume-limited survey.  These two scenarios, flux-limited/spatially uniform and volume-limited, almost certainly bracket the truth.  The rest of our calculations use an average or composite of these two limiting cases.

Assuming a flux-limited survey, the total gamma-ray flux from {\it Fermi}-resolved MSPs in a luminosity range $[L,\,L+\delta L]$ is
\begin{equation}
F_{\rm tot} \propto \int_0^{r_{\rm max}} dr\,r^2 \frac{L}{r^2} \frac{dN}{dL} \delta L
= r_{\rm max} L \frac{dN}{dL} \delta L~,
\label{eq:Ftot}
\end{equation}
where $dN/dL$ is the differential number density and $r_{\rm max} \propto \sqrt{L/F_{\rm min}}$ is the maximum radius out to which the source could be detected.  For a source population around the Galactic center, we wish to know the total luminosity per unit volume, which is simply given by
\begin{equation}
F_{\rm GC} \propto \frac{L}{r_{\rm GC}^2} \frac{dN}{dL} \delta L~.
\label{eq:Ltot}
\end{equation}
We combine Equations \eqref{eq:Ftot} and \eqref{eq:Ltot} to obtain
\begin{equation}
F_{\rm GC} \propto F_{\rm tot} L^{-1/2}~. 
\end{equation}
The integrated spectrum from an unresolved MSP population near the Galactic center may therefore be estimated by scaling the observed flux density of each object by that object's luminosity to the $-1/2$ power, and simply adding all of the scaled flux densities together.

\begin{figure}
\centering
\includegraphics[width=\linewidth]{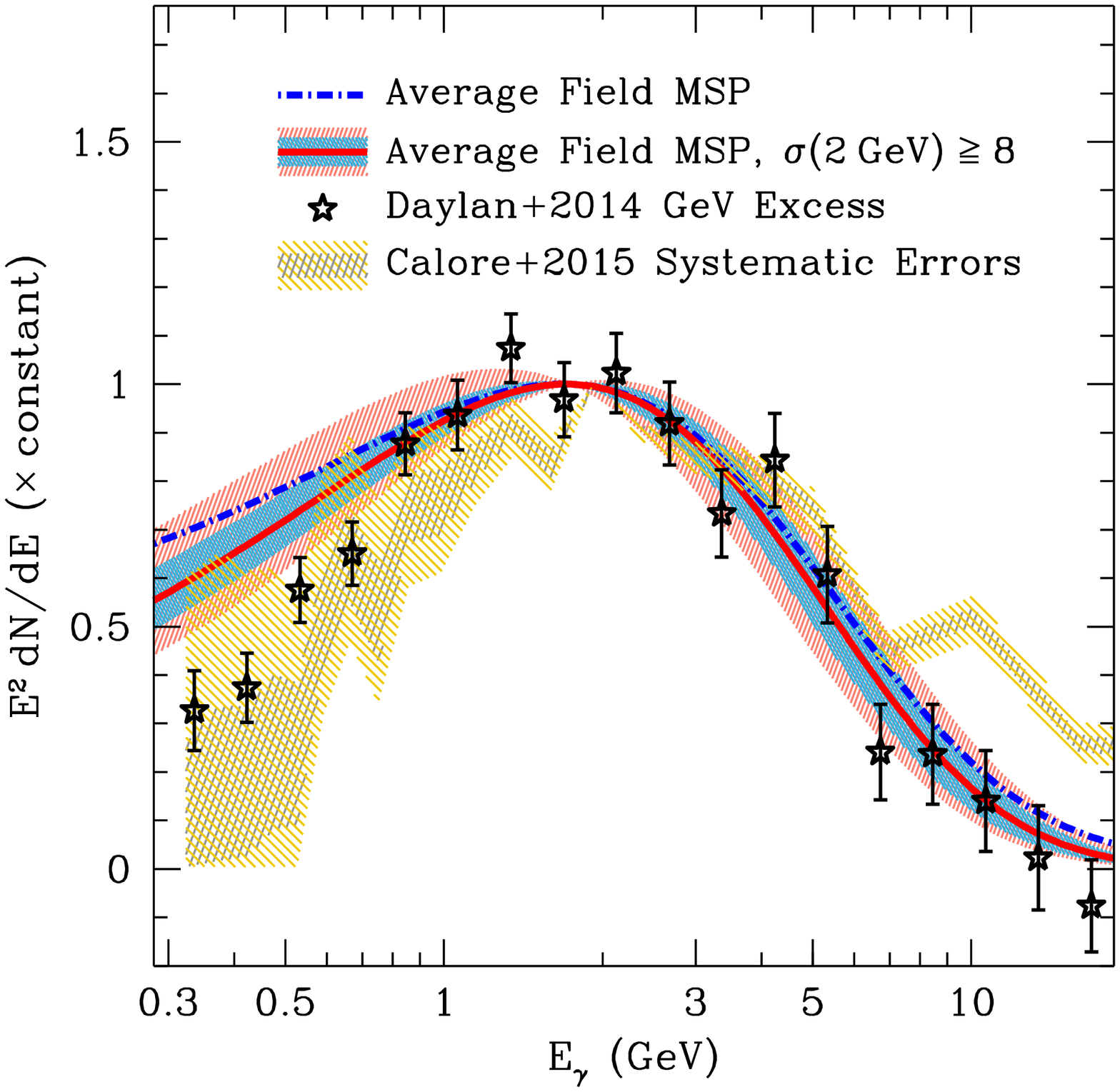}
\caption{The average spectrum of {\it Fermi}-detected field MSPs adopting the fitted spectral parameters of \cite{Cholis+Hooper+Linden_2014}.  The dotted-dashed blue line is the unweighted average spectrum.  The red line has selected only those MSPs detectable based only on their 1--3 GeV flux (45 of 59 MSPs), and is the average of the spectra expected for a population at uniform distance assuming the \cite{Cholis+Hooper+Linden_2014} to be volume-limited and flux-limited.  These scenarios almost certainly bracket the truth.  The blue and orange hatching show 1$\sigma$ and 2$\sigma$ sample variances as estimated using bootstrap resampling.  We have neglected errors in the MSP distances and in the spectral measurements; both would tend to alleviate the discrepancy with the observed Galactic center excess \citep{Daylan+Finkbeiner+Hooper+etal_2014}.  The error bars on the \cite{Daylan+Finkbeiner+Hooper+etal_2014} fits are only statistical; systematic errors (which are spectrally correlated) are neglected.  The gold and gray hatching show 1$\sigma$ and 2$\sigma$ systematic uncertainties (neglecting statistical errors) as estimated by \cite{Calore+Cholis+Weniger_2015}. 
}
\label{fig:spec}
\end{figure}

Figure \ref{fig:spec} shows the results, with all spectra normalized to their 1.7 GeV values.  The blue dotted-dashed line shows the result for simply adding together all MSP spectra without selecting them by 1--3 GeV flux and without scaling by luminosity.  The red line is the average of the spectra with and without weighting by $L^{-1/2}$, i.e, assuming volume-limited and flux-limited samples, respectively.  The blue and orange hatching show the 1$\sigma$ and 2$\sigma$ uncertainties in the red spectrum as estimated from bootstrap resampling of the 45 MSPs.  For this exercise, we have adopted the fitted spectra in Table I of \cite{Cholis+Hooper+Linden_2014} and have neglected measurement errors, fitting errors, and distance errors.  

The difference in Figure \ref{fig:spec} between the scaled and unscaled spectra results from a correlation between luminosity and spectral index.  Distance errors will tend to blur this correlation; the MSP spectrum of a population at a single distance is likely to be slightly harder than the red line in Figure \ref{fig:spec}.  Including this effect and adding measurement errors would not bring the MSP spectrum into perfect agreement with the Galactic center excess, but it could bring the 1$\sigma$ discrepancy to as little as $\sim$20--30\% at 500 MeV.  Selecting only those MSPs with $|b|>10^\circ$ (38 of the 45 that pass our 1--3 GeV signal-to-noise cut) would also marginally improve the agreement with the spectrum of the GeV excess.  

The discrepancy between our estimated average MSP spectrum and the GeV excess is only significant at the lowest energies ($<$800 MeV) where {\it Fermi}'s sensitivity is rapidly falling.  Uncertainties in Galactic diffuse emission are largest here \citep{Calore+Cholis+Weniger_2015}.  As a result, there are spectrally correlated systematic errors in the spectrum of the GeV excess not shown in the black stars of Figure \ref{fig:spec}.  Systematic errors can be quite large, and can also arise from the method of masking point sources and from the assumed morphology of the excess, among other aspects of the fitting \citep{Daylan+Finkbeiner+Hooper+etal_2014, Calore+Cholis+Weniger_2015}.  Figure \ref{fig:spec} also shows the systematic errors from varying the diffuse backgrounds as estimated by \cite{Calore+Cholis+Weniger_2015}.  These gray and gold hatched regions neglect statistical errors.

\section{Prospects for Radio Detections} \label{sec:discussion}

Our results show that a population of disrupted globular clusters, which must exist to explain the current clusters, naturally predicts a field population of MSPs in the Galaxy's inner few kpc.  These MSPs satisfy the spatial, spectral, and luminosity requirements imposed by the {\it Fermi} observations.  A large population of MSPs in a nuclear star cluster is another necessary consequence of a population of disrupted massive globular clusters.   Such a population explains the 20--40 keV X-ray emission seen by {\it NuSTAR} \citep{Perez+Hailey+Bauer+etal_2015} and implies that many of the unidentified {\it Chandra} point sources may be MSPs \citep{Muno+Baganoff+Bautz+etal_2004,Perez+Hailey+Bauer+etal_2015}.  {\it Astro-H} \citep{Takahashi+Mitsuda+Kelley+etal_2010} will also be sensitive to high-energy X-rays, and could confirm the {\it NuSTAR} results.  A population of $\sim$1000 MSPs around Sgr A* can also explain the observed TeV emission by inverse Compton scattering of the dense interstellar radiation field \citep{Bednarek+Sobczak_2013}.  

Radio observations could individually detect our predicted MSPs and confirm their identities.  
However, the bulk of the radio observations to date have focused not on scales of tens to thousands of pc, where most of our predicted MSPs lie, but in the innermost pc.
This was motivated by theoretical estimates predicting $\sim$100--1000 pulsars formed in situ within 0.02 pc of Sgr A* \citep{Pfahl+Loeb_2004}.  More recently, \cite{Faucher-Giguere+Loeb_2011} noted that the encounter rate in the inner 1 pc of the central star cluster is comparable to that of the globular cluster Terzan 5 (which has many MSPs), and estimated that up to $\sim$1200 MSPs may be present in this region due to the deeper gravitational potential well of Sgr A*.  The disrupted globular cluster scenario instead predicts these MSPs to be found over a larger region: we predict $\sim$1,000 MSPs within 3 pc of Sgr A*, and a further $\sim$1,000 MSPs within 300 pc ($2^\circ$, see Figure \ref{fig:integral_GCE}).  

MSP observations towards the Galactic center are extremely challenging because of the large dispersion measures.  Radio pulses at a frequency $\nu$ are broadened by an amount $\tau = (1.3 \pm 0.2) (\nu/{\rm GHz})^{-3.8\pm 0.2}$ \citep[with $\tau$ in seconds, ][]{2014ApJ...780L...3S}, implying that MSPs may not be observed below $\sim$8 GHz. The radio intensity of pulsars scales steeply with frequency ($I\propto \nu^{-1.6}$ to $\nu^{-1.8}$, \citealt{Kramer+etal_1998}), so high-frequency detections require extended integration times. 

While discovering and timing MSPs 0.001 pc from the central supermassive black hole would offer tantalizing measurements of general relativity and tests of alternative theories of gravity \citep{Wex+Kopeikin_1999,Kramer+etal_2004,Cordes+etal_2004,Pfahl+Loeb_2004,Liu+etal_2012}, discovering MSPs further out within 10 pc would also be invaluable.  Such MSPs could be used to measure the properties of the nuclear star cluster, find intermediate mass black holes, and measure the gravitational waves of the Galactic center \citep{Kocsis+Ray+PortegiesZwart_2012}. If the nuclear star cluster indeed formed from disrupted globular clusters, we predict $\sim$1000 MSPs within $\sim$10 pc of the Galactic center; such a population of MSPs can account for the unresolved {\it Fermi} flux seen by \cite{Abazajian+Canac+Horiuchi+etal_2014}.  Future high frequency radio surveys will have the frequency coverage and sensitivity needed to detect this MSP population \citep{Chennamangala+Lorimer_2014,Macquart+Kanekar_2015}. Even larger radio surveys such as the square kilometer array (SKA) on 100 pc to 2 kpc scales are required to confirm or disprove the disrupted globular cluster origin of MSPs in the Galactic bulge.

\section{Conclusions} \label{sec:conclusions}

The {\it Fermi} Galactic center excess is in excellent agreement with independent predictions of the population of MSPs produced in disrupted globular clusters.  This astrophysical model appears to fit the observations as well as dark matter annihilation, but without any free parameters.  MSPs from disrupted clusters also provide an excellent match to the observed emission near Sgr A* from hard X-rays through very hard gamma rays.  If the bulge indeed contains a large population of stars from long-dead clusters, such MSPs form a background that must necessarily be present in the {\it Fermi} data.  

The observed emission extends at least $\sim$2 kpc from the Galactic center 
\citep{Hooper+Slatyer_2013}, far from the nuclear star cluster around Sgr A* where dynamical formation of MSPs is plausible. LMXBs burn out after the disruption of globular clusters, reducing their relative numbers in the galactic bulge, consistent with the lack of LMXB observations  \citep[c.f.][]{Cholis+Hooper+Linden_2015}.
We conclude that the dominant MSP population is not likely to have formed under the current conditions in the bulge, but was deposited by dissolving globular clusters.  If the {\it Fermi} excess is indeed the relic of a previous large population of globular clusters, it provides the first direct evidence for their existence,
and strongly supports the theory for the globular cluster origin of the nuclear star cluster. Future radio observations may be directly sensitive to these MSPs and could offer decisive evidence of a broad distribution of MSPs deposited by globular clusters.

While our results disfavor a dark matter interpretation of the GeV excess, they show that {\it Fermi} can offer a new probe of the formation history of the bulge, and of the evolution of the Galaxy's globular cluster system.  Our reevaluation of field MSP luminosities, combined with the results of \cite{Lee+Lisanti+Safdi+etal_2015} and \cite{Bartels+Krishnamurthy+Weniger_2015}, suggest that we will soon begin to resolve the brightest of these fossils.

\acknowledgments{The authors thank Scott Tremaine, Doron Kushnir, Mariangela Lisanti, Neal Dalal, and Oleg Gnedin for very useful discussions, and an anonymous referee for helpful suggestions.  B.K.~gratefully acknowledges support from the W.M.~Keck Foundation Fund of the Institute for Advanced Study, NASA grants NNX11AF29G and 13-ATP13-0056, and NSF grant AST-1406166. This work was performed in part under contract with the Jet Propulsion Laboratory (JPL) funded by NASA through the Sagan Fellowship Program executed by the NASA Exoplanet Science Institute. This work was supported in part by the European Research Council under the European Union\textsc{\char13}s Horizon 2020 Programme, ERC-2014-STG grant GalNUC 638435.
}

\bibliographystyle{apj_eprint}
\bibliography{refs}

\appendix

\section{The evolution of a population of globular clusters} \label{sec:clusterpop}

We adopt the semianalytical method of  \citet{Gnedin+Ostriker+Tremaine_2014} to calculate the evolution of a population of globular clusters in the Milky Way. The only result we use is the mass distribution of disrupted clusters, which we take directly from their Figure 3. We provide a brief summary of the basic assumptions of the \citet{Gnedin+Ostriker+Tremaine_2014} model here, and refer the reader to that paper for details. Using parameters that match the observed properties of young star clusters, the model recovers the present-day observed masses and radial distribution of globular clusters, and it predicts the total radial mass distribution of globular cluster mass deposited in the Galactic bulge. The initial mass distribution of globular clusters is set to $dN/dM\propto M^{-2}$ between $10^4$--$10^7\,M_{\odot}$, and the initial radial distribution of GCs is set to follow the mass distribution of Galactic field stars scaled down uniformly to $1.2\%$ assuming that all globular clusters formed at $z=3$. This distribution places equal mass in logarithmic bins, i.e., there is the same amount of stellar mass in $10^6$--$10^7$ $M_\odot$ globular clusters as in $10^4$--$10^5$ $M_\odot$ clusters.  All components (globular clusters, field stars and dark matter) are spherically distributed around the Galactic center; the field stars follow a Sersic profile with an index $n_s=2.2$, the enclosed mass is approximately $M(R) \sim 10^5\,M_{\odot} (R/100\,{\rm pc})^{2.4}$ if $R\lesssim 1$ kpc; the dark matter follows an NFW profile $\rho_{\rm dm}\propto (R/R_s)^{-1}(1+R/R_s)^{-2}$ where $R_s=20$ kpc and a total mass within $12 R_s$ is $10^{12}\,M_{\odot}$  \citep{Navarro+Frenk+White_1997}; and there is a supermassive black hole of $4\times 10^6\,M_{\odot}$ at the center. 

Once the model is initialized, the clusters move on circular orbits in the instantaneous gravitational field of dark matter, stars, globular clusters, and the deposited mass from globular clusters, and they spiral inwards due to dynamical friction. The inspiral time is proportional to $M^{-1}$ \citep{Binney+Tremaine_2008}, implying that more massive clusters segregate inwards more quickly. 
The globular clusters evolve due to mass loss through stellar evolution, they slowly evaporate independently of their location relative to the galactic center, and they lose mass through tidal stripping by the galaxy. The orbital time, dynamical friction time, isolated cluster evaporation time, and tidal disruption time are given, respectively, as
\begin{align}\label{eq:torb}
t_{\rm orb} & = 0.06\,{\rm Gyr}\,\frac{R}{\rm kpc} \left( \frac{v_{c}(R,M)}{100\,{\rm km\,s^{-1}}}\right)^{-1} \\\label{eq:t_df}
t_{\rm df} &= 45\,{\rm Gyr}\,\left(\frac{R}{\rm kpc}\right)^{2} \left( \frac{v_c(R,M)}{100 {\rm km\, s^{-1}}} \right)^{-1} \left(\frac{m}{10^5\,M_{\odot}}\right)^{-1}f_\epsilon\\\label{eq:t_iso}
t_{\rm iso} &= 17 \,{\rm Gyr}\,\frac{m}{2\times 10^5\,M_{\odot}}\\\label{eq:t_tide}
t_{\rm tid} &= 67\,\left(\frac{m}{2\times 10^5\,M_{\odot}}\right)^{\alpha} t_{\rm orb}(R,M)
\end{align}
where $R$ is the orbital radius, $m$ is the cluster mass, $M$ is the enclosed mass, $\alpha=2/3$ and $f_\epsilon=0.5$, $v_c(R,M)=(GM/R)^{1/2}$ is the circular velocity. 
The clusters are followed individually and modeled using their average properties: half-mass radius, total mass, and average density, as
\begin{align}
\frac{dm}{dt} &= - \frac{m}{\min(t_{\rm tid}, t_{\rm iso}, t_{\rm wind}) }
\\
\frac{d R^2}{d t} &= - \frac{R^2}{t_{\rm df}}
\end{align}
Here $t_{\rm wind}$ is the timescale on which mass is lost due to stellar evolution and winds following \citep{Prieto+Gnedin_2008}. Approximately $30\%$ of mass is lost during the first 0.3 Gyr, and another $10\%$ during the following 10 Gyr. 
The circular velocity and enclosed mass are updated as the radial mass distribution changes during the inward migration and evaporation of globular clusters.

The average density of the cluster is assumed to vary with mass as \citep{Gnedin+Ostriker+Tremaine_2014}: 
\begin{align}\label{e:rhoh}
\rho_h &= 10^3\,\frac{M_{\odot}}{{\rm pc}^3} \times \min\left\{10^2, \max\left[ 1, \left(\frac{m}{10^5\,M_{\odot}}\right)^2\right]\right\}\\
r_h &= \left( \frac{ 3 m }{ 8 \pi \rho_h(m) } \right)^{1/3}
\end{align}
A globular cluster is disrupted when the mean enclosed density in dark matter, gas, and stars exceeds the average density of the cluster itself,
\begin{equation}\label{e:rhodisrupt}
\rho_h < \frac{\left( v_c(R, M) \right)^2}{2\pi G R^2}.
\end{equation}
Heavy, dense clusters can sink closer to the center before getting disrupted. The mass weighted mean lifetime of disrupted clusters is typically several Gyr. Thus, an average disrupted cluster may have had a similar number of MSPs per unit mass as similar mass clusters further out which survived disruption until the present (see Section \ref{sec:lumscale}). 

In this model, the surviving globular clusters have an approximately lognormal mass distribution and a nonuniform radial distribution that is consistent with observations. The globular clusters that do not survive are typically disrupted before they reach the very center of the Galaxy, creating a characteristic cored density profile. The mass of the disrupted globular clusters exceeds the initial stellar mass in the the nuclear star cluster, the very central region of the galactic bulge, delivering a few $10^7$ $M_\odot$ within $\sim$10 pc of Sgr A*.  An additional $\sim$10$^8$ $M_\odot$ is deposited interior to $\sim$1 kpc (see Figure 3 in \citealt{Gnedin+Ostriker+Tremaine_2014}). The mass from disrupted clusters is deposited roughly spherically with a density decreasing with radius approximately as $\rho \sim r^{-2.2}$ at 1 kpc.
Here, the exponent depends on the assumed details of the initial cluster population, but is roughly constant (within a few tenths) between $\sim$200 pc and a few kpc ($\sim$1$^\circ$ and 20$^\circ$ projected at 8.3 kpc).  

While this toy model captures many of the essential features of globular cluster evolution within galaxies, it neglects several possibly important details. These include core collapse; binary and multibody interactions which may heat the cluster or eject stars; gas effects (accretion, inflow, star formation); resonant interactions, violent relaxation, radiative or thermal feedback from supernova explosions or an active galactic nucleus; the effects of galactic anisotropy (disk, bar, spiral arms); the effect of tidal shocks when crossing vertically through the galactic disk; the collision of globular clusters; the formation of new star clusters; the effects of galaxy mergers; and supermassive black hole binaries.  Some clusters do display indications of interesting formation histories that are hinted at by these complicating effects but are not captured in the toy model \citep{Bedin+Piotto+Anderson+etal_2004, Marino+Villanova+Piotto+etal_2008, Ferraro+Dalessandro+Mucciarelli+etal_2009, Marino+Milone+Piotto+etal_2009, Yong+Roedere+Grundahl+etal_2014, Milone+Marino+Piotto+etal_2015}. 
Some of these effects may have an influence on the radial mass-loss profile, but most will not affect the predicted spherical morphology of the tidal debris, as long as the initial distribution of globular clusters is roughly spherical and the rate of mass loss is slow over an orbital time. 
For example, a possible source of asphericity may result from tidal shocks generated by the cluster crossing the Galactic disk, which catalyses evaporation or core collapse. However, the characteristic timescale of tidal shocks ranges between 3 to 10 half-mass relaxation times \citep{Gnedin+Lee+Ostriker_1999}, which is between 0.1 and several Gyr, longer than the orbital time within 2 kpc (see Eq.~\ref{eq:torb}). The tidal debris from globular clusters overlapping with the gamma-ray excess may remain spherical in a wider class of models than \citet{Gnedin+Ostriker+Tremaine_2014}.

\end{document}